\documentclass[manuscript,screen]{acmart}

\AtBeginDocument{%
  \providecommand\BibTeX{{%
    \normalfont B\kern-0.5em{\scshape i\kern-0.25em b}\kern-0.8em\TeX}}}

\setcopyright{iw3c2w3g}
\copyrightyear{2023}
\acmYear{2023}
\acmDOI{XXXXXXX.XXXXXXX}

\begin{document}

\title{Approaching Unanticipated Consequences}


\author{Andrew Darby}\email{a.darby@lancaster.ac.uk}\affiliation{\institution{Lancaster University}\country{UK}}


\author{Pete Sawyer}\email{p.sawyer@aston.ac.uk}\affiliation{\institution{Aston University}\country{UK}}

\author{Nelly Bencomo}\email{nelly.bencomo@durham.ac.uk}\affiliation{\institution{Durham University}\country{UK}}


\renewcommand{\shortauthors}{Darby et al.}

\begin{abstract}
In an ever-changing world, even software that fulfils its requirements may have un-envisioned aftereffects with significant impacts. We explored how such impacts can be better understood at the pre-design phase in support of organisational preparedness. We considered three real-world case studies and engaged with literature from several disciplines to develop a conceptual framework. Across three workshops with industry practitioners and academics, creative strategies, from speculative design practices were used to prompt engagement with the framework. We found participant groups navigated the model with either a convergent or divergent intent. The academics, operating in an exploratory mode, came to a broad understanding of a class of technologies through its impacts. Operating in 
an anticipatory
mode the industry practitioners came to a specific understanding of a technology’s potential in their workplace. The study demonstrated potential for the conceptual framework to be used as a tool with implications for research and practice. \end{abstract}



\maketitle

\section[Introduction]{Introduction}\label{Introduction}

We live in a rapidly-changing world in which some of the primary motors of change are innovations in software. Some of these innovations change how we do things in an incremental way; others radically disrupt our personal, social and business lives. Most of these changes are intentional, planned effects of innovation, but others are unforeseen, emergent properties. In the latter category, examples include the amplification of extreme views and the facilitation of user surveillance afforded by social networking, the reinforcement of social injustices caused by facial recognition systems, and the contribution to global heating by cryptocurrency mining. Unanticipated real-world impacts present a problem that is becoming increasingly urgent as more and more decisions are devolved to the opaque workings of machine learning technologies.

Historically there have been a number of reasons for a lack of attention to the consequences of innovation, \citeauthor{rogers2010diffusion} enumerates them, noting that
`Consequences have not been studied adequately because (1) change agencies have overemphasized adoption per se, assuming that the consequences will be positive; (2) the usual survey research methods may be inappropriate for investigating consequences; and (3) consequences are difficult to measure' \citep{rogers2010diffusion}.

Consequences, explained in relation to diffusion theory, are 
`the changes that occur to an individual or to a social system as a result of the adoption or rejection of an innovation' \citep{rogers2010diffusion}. The dimensions of consequence may be understood in relation to the proximity of a consequence to the design decision (direct vs indirect), the desirability of the consequences arising from a design decision (desirable vs. undesirable) and the predictability of the consequence (anticipated vs. unanticipated) 
\citep{rogers2010diffusion}. 

Technological innovation, such as developing new software products, deals in intended consequences that are predictable, desirable and direct. 
Broadly, change agents want their innovations to have beneficial consequences. However, potential \textit{unanticipated consequences}, 
`changes that are neither intended nor recognized by the members of a system' \citep{rogers2010diffusion}, may still emerge. We believe that 
software designers (broadly defined to include requirements engineers) should be concerned about their innovations' societal, business and environmental side-effects, but there are challenges. In particular, there are a lack of `reliable methods for anticipating undesirable consequences of innovation' \citep{sveiby2009unintended}. So, even where there exists a motivation to investigate the wider impacts of innovation, the tools to do so barely exist within software development. 
This is the problem that we are interested in and in this paper, we start from the position that any such tools are going to have to direct 
software designers' creativity to look beyond the presumed benefits that traditionally drive software requirements and design decisions, towards considering how their potential realisations might impact the world, and potential futures, more widely. 


\section[Related Work]{Related Work}\label{Related Work}

Below we note the various strands of research that influence our work, including; Requirements Engineering, HCI, Design Thinking, Speculative Design and Futures acknowledging their different relationships with the future. These disciplines operationalise futures-orientated work in different ways that span from product-specific probable futures that are the nearest that they can possibly be to coming into existence to the broadest stroke articulations of societal futures that are almost entirely unlikely. Below, we note how the various disciplines' focuses and their objects of study relate to the potentiality and proximity of their imagined futures. 

Successful software requires its designer to discover features and properties of the software that will satisfy the needs and desires of the software's customers and users, and subject to constraints such as cost, standards, compatability, testability and so on. These are software requirements. \citeauthor{maidenRE2010} argue that 'By framing requirements engineering as creative problem solving we can gain new insights into it, and recruit new knowledge from other disciplines to understand it better and support it more effectively' \citep{maidenRE2010}. 



Part of the challenge addressed by \citeauthor{maidenRE2010} is that developing `good' software requirements cannot be done without a detailed understanding of the (business/market/social/physical/political/etc.) world in which the software will operate and on which it will act. 
The world is complex and, crucially, not static. Requirements engineers, software designers, usability experts and everyone involved in the conception and development of software make decisions informed in part by assumed futures; their conception of what the world will look like during the lifetime of their software. Making the assumed future explicit is rare in software development practice, but explicitly envisioning the future has found utility in HCI and Ubiquitous Computing research \citep{reeves2012envisioning}.



Orientated towards the commercially-viable probable future, Design Thinking `is a discipline that uses the designer’s sensibility and methods to match people’s needs with what is technologically feasible and what a viable business strategy can convert into customer value and market opportunity' \citep{brown2008}. It has been configured as a methods pack and used to address `open, complex problems' in prospective design practices and crucially to challenge the assumptions within initial briefs \citep{dorst2011}. To do so, it makes particular use of abductive reasoning strategies, and the development of frames, where `a ``frame'' is the general implication that by applying a certain working principle we will create a specific value' \citep{dorst2011}.

Alternative design practices, like Speculative Design, set aside the constraints of commercial imperatives in order to critically explore the conjunction of desirability and potentiality \citep{auger2013,dunne2013} querying ideologies and values on the way. Methods such as design fiction \citep{Bleecker2009} which problematize and borrow from Design Thinking among a wide range of other creative practices, are used to focus on the point at which the plausible is on the cusp of becoming probable. The work is underpinned by a focus on scientific possibility and believability, as well as the proximity of the articulated future to the present.
Design fiction may be understood as `the practice of creating tangible and evocative prototypes from possible near futures, to help discover and represent the consequences of decision making' \citep{manualDF}.

As a whole the futures field tries to  `provide policy-makers and others with views, images, alternatives etc. about futures in order to inform the present' \citep{slaughter2}. The future is fiercely contested, no two people make sense of the world in exactly the same way and `different worldviews and values disclose different truths' \citep{slaughter2} leading to myriad futures. To help navigate the nature and complexity of futures, \citeauthor{voros2001} helpfully breaks them down as to the potentiality and desirability of them coming to fruition –– marking them as probable, plausible, possible and preferable futures \citep{voros2001}. While the futures field takes a broad view of, and interest in, possible futures, there is 
a particular focus on plausible futures in service to organisational preparedness. 

Our work draws on aspects of these various approaches. We seek a practicable means for designers to reason about the future beyond the merely probable, assumed futures that drive many design decisions to enable decisions that are more robust when the probable does not materialise.

\section[Research Design]{Research Design}\label{Methods}

This section offers a brief overview of the research design, before noting our approach, recruitment methods and remuneration strategies and analysis.

The research study was conducted in four phases. In the first we made reflections on technologies in the world making use of three case studies. In the second phase we brought those reflections into conversation with concepts and theories from the various literatures noted in section \ref{Related Work} and developed a visualisation of a conceptual framework. In the third, that model was used to guide participant enquiry in three workshops, two with industry professionals and another with academics, and in the fourth phase we analysed the utility of the model in use.



As the unanticipated consequences of technologies were revealed in national UK news features we engaged in an extended dialogue around these instances. Such fortuitous case studies allowed us to explore unanticipated consequence in the actual world as part of three different software or software-intensive products. Specifically, we looked at the UK Post Office's Horizon electronic point-of-sale system, see \ref{Horizon}, DJI's commercial drone technology in Ukraine, see \ref{Ukraine}, and Apple's tracker technology, AirTags, see \ref{AirTag}.

We drew on the case studies and the the literature to synthesise a conceptual framework. This was realised as a visual model for use in a workshop setting, to help explain the workshop activities to participants and to orient both participants and facilitators during the workshops.


The workshops took place at the participant group's organisational sites in standard meeting rooms. The first and second workshop were held on 28/06/22 and 04/10/22 and the third workshop on the 11/10/22. The lead author facilitated the workshop in-person for all three workshops, while the second author acted as an in-person observer for the first and third workshop and observed using teams in the second workshop, their field notes provided a rich description of events for subsequent analysis.

The research team purposively selected the participant groups though individuals' participation was voluntary. Participants in the first and second workshop comprised practitioners from our industry partner, a large telecomms provider.


The participants in the third workshop included academics and postgraduates drawn from a university Computer Science department in the North East of England. Each of the workshops lasted two hours and the postgraduate student participants were offered a nominal remuneration --a £25 Amazon voucher-- to thank them for their participation.


Following the workshops, we reflected on what the participants achieved relative to our goals and identified what  worked well and what worked less well.


\section[Findings]{Findings}\label{Findings}

This section describes the case studies and what we learned form them through the lens of consequences and values. We briefly outline the literature and describe the conceptual framework. The section concludes with a commentary on the workshops.


\subsection[Case studies - Reflecting on the world]{Case studies - Reflecting on the world}\label{Findings1}

In the following three subsections, we discuss examples of recent innovations in which insufficient attention appears to have been paid to plausible consequences and which we believe illustrate the need for developers to have better tools with which to explore the plausible futures around technological innovations. We highlight the UK Post Office's Horizon electronic point-of-sale system, DJI's commercial drone technology in Ukraine, and Apple's tracker technology, AirTags. As we discuss these instances we highlight key concepts, in parentheses, that will be taken up and made clear in section \ref{Findings2}.

\subsubsection[The Post Office Horizon system
]{The Post Office Horizon system
}\label{Horizon}

The Horizon electronic point of sale (PoS) and accounting system, `Europe's largest non-military IT contract' \citep[Technical Appendix 15]{bates3408}, was developed by Fujitsu for the UK Post Office. Subpostmasters, self-employed agents running post offices, were contractually required to use Horizon. Inevitably, Horizon had defects, some of which are now known to have caused accounting errors. The service contracts between Fujitsu and the Post Office disincentivised both parties from believing or acting on subpostmasters' reports of Horizon's failures. Instead, the Post Office's policy was to prosecute any subpostmasters who did not make good any shortfall from their own funds, deeming them liable for the errors \citep{bates606}. No-one in the Post Office, Fujitsu or the courts seemed to consider it suspicious that a large body of people of previously good character had suddenly taken to embezzlement, and that this conversion to criminality coincided exactly with the introduction of a new PoS and accounting system. As a result, hundreds of Subpostmasters suffered financial penalties. Some were jailed. Some took their own lives. Eventually, a group action by 550 subpostmasters exposed Horizon's defects and the egregiousness of the Post Office and Fujitsu's corporate behaviour \citep{bbc, wallis21}.

Horizon exhibited \emph{unanticipated negative consequences}. Every non-trivial software system has defects and  these sometimes result in failures, such as producing the accounting errors produced by Horizon. However, these need not have produced what were the real negative consequences; those on subpostmasters' livelihoods, health and reputations. It was the Post Office's corporate behaviour, in conjunction with a myopic refusal on the part of the Post Office and Fujitsu to take action despite the evidence of defects, that was responsible for these. 

\subsubsection[The Apple AirTag]{The Apple AirTag}\label{AirTag}

Marketed as a way to keep track of your things the Apple AirTag, released in 2021, was predictably and quickly misused, e.g. for stalking and car theft. Non-Apple, primarily Android, phone users are particularly vulnerable to AirTag misuse. Apple eventually responded to Android phone users' concerns with the release of the Tracker Detect app. Even now, however, while the Apple ecosystem automatically alerts iPhone users to Airtags in their vicinity, Android users have to actively run scans for AirTags. This is despite TU Darmstadt's  AirGuard app \citep{airtags} demonstrating that generating automatic alerts for Android users was possible.

The potential indirect collateral damage to Non-Apple users arising from Apple's primary design decisions has either been deemed an \emph{acceptable negative} or was an \emph{unanticipated negative consequence}. It can be argued that AirTags embody values that include \emph{wealth} (iPhones are typically at the more expensive end of the market) and selectively prioritise other values, most obviously users' personal \emph{security}.

\subsubsection[The war in Ukraine and DJI Drones]{The war in Ukraine and DJI Drones}\label{Ukraine}

 The use of commercial drone technology in the Ukraine War has been of particular interest to US policymakers in terms of near-term plausible futures \citep{droningon} and to the Ukrainian government in terms of the direct consequences surrounding their use in the war. Both Ukraine and Russia have used commercial drones to carry out reconnaissance. Ukraine has also modified commercial drones to drop artillery. Russia has targeted drone pilots with artillery bombardments using DJI's \emph{Aeroscope} drone detection product. Whether these repurposings of the drones and Aeroscope were \emph{anticipated} or \emph{unanticipated consequences} to their manufacturers is unclear. 
 
 Ukraine called on the commercial drone manafacturer, DJI, to block Russia's use of their drones in the region \citep{drones}. DJI's CEO responded stating that the company could not change the product for a number of technical reasons \citep{DroneXL}. Without appropriate technical solutions to navigate the demands of the two sales territories, the company stopped sales in Russia and Ukraine \citep{vice}. The company's drones still operate in both regions. It should be noted that DJI is a Chinese company and may be operating under governmental influence or even direction in this matter. 
 

Insights regarding the cases described in section \ref{Horizon}, section \ref{AirTag} and section \ref{Ukraine} were used drive the development of the conceptual framework explicated below.

\subsection[Engage with literature and develop a conceptual framework]{Engage with literature and develop a conceptual framework}\label{Findings2}


\begin{figure}[htbp!] \centering\setlength{\fboxrule}{0.25pt} \setlength{\fboxsep}{0pt}\fbox{\includegraphics[height=.6\textwidth]{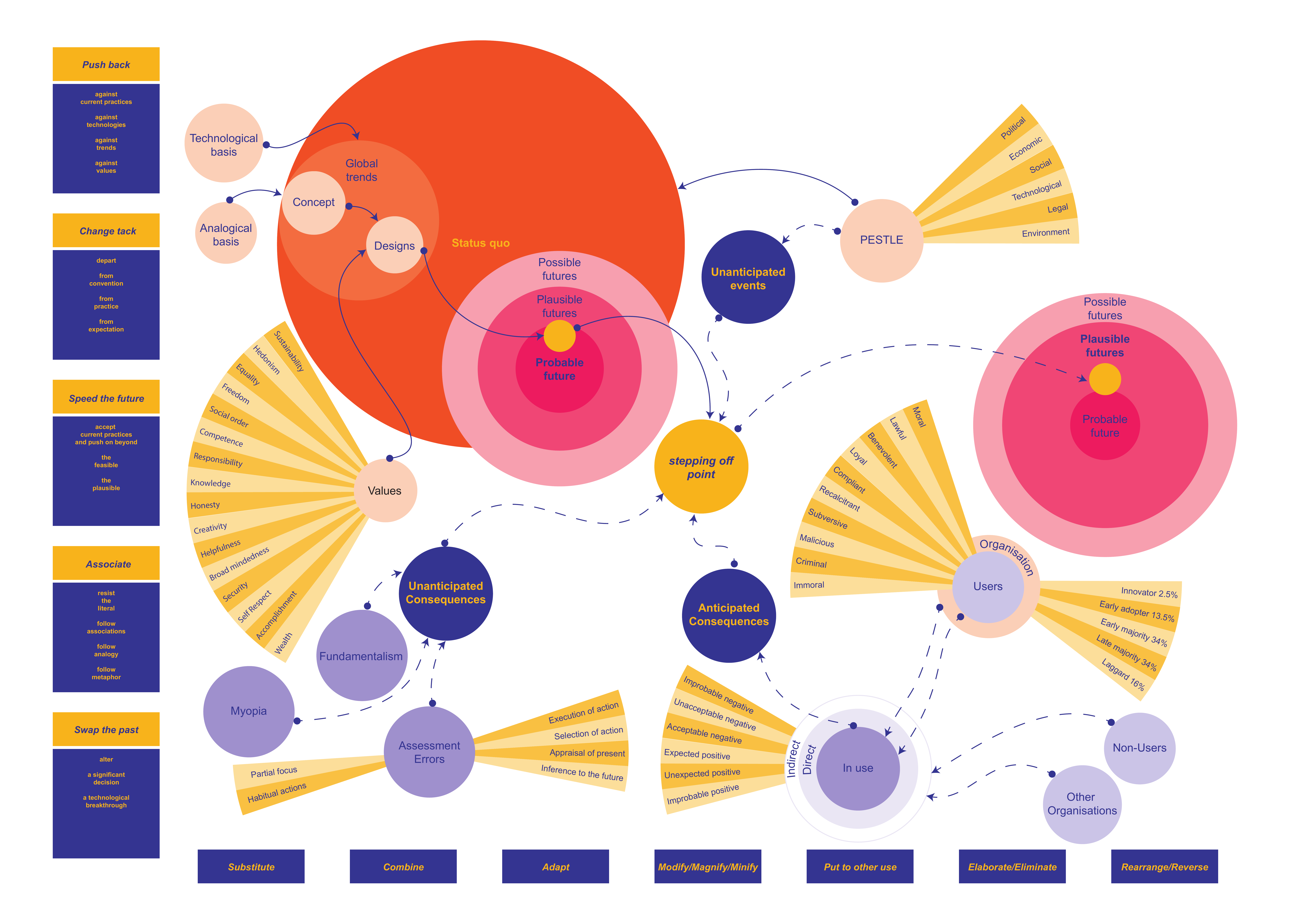}}\caption[Consequences and Futures Model]{Consequences and Futures Model} \label{fig:Consequences and Futures model}\end{figure}

We developed the \textit{Consequences and Futures Model}, see figure \ref{fig:Consequences and Futures model}, described below in response to the reflections discussed in section \ref{Findings1} 
The conceptual framework is split into four main areas; `Status Quo', `Unanticipated Events', `Anticipated Consequences'
and `Unanticipated Consequences'. A stack of circles representing the reducing potentiality of \textit{Possible, Plausible and Probable} futures
\citep{voros2001} appear twice, with the first instance overlaying the `Status Quo' giving focus to the \textit{Probable} and the second set away to one side highlighting the \textit{Plausible}. To support designers' (anyone participating in a design or requirements exercise, software practitioner, customer, user, etc.) engagement with the conceptual framework five speculative strategies drawn from \citep{pierce2021} are presented alongside a cognitive heuristic \citep{eberle}.

\subsubsection[Status Quo]{Status Quo}\label{Status Quo}
This part of the model invites the user to consider the \textit{Analogical} \citep{maidenRE2010},
and \textit{Technological basis} of a design \textit{Concept} and to further consider how that \textit{Concept} and the \textit{Design}(s) that springs from it form part of \textit{Global trends}. It asks the designer to consider how \textit{Values} are embedded in a design, how a design acts prospectively to describe a probable future, and how the conditions of the wider world support that possibility.
The framework adopts a values taxonomy \cite{Sutcliffe} to enable speculation regarding values embedded in the design. The values can be employed to query organisational \textit{Myopia} and \textit{Fundamentalism}, in \ref{Unanticipated Consequences}, as well as supporting exploration of \textit{Designers'} attitudes, in \ref{Anticipated Consequences}.


\subsubsection[Unanticipated Events]{Unanticipated Events}\label{Unanticipated Events}

As they develop an innovation companies plan for the market to be configured in a particular manner. After launch an innovation’s success is reliant, to some degree, on the stability of its context of use. However, in the real world 
contexts of operation are unpredictable and subject to change, which may have radical implications for an innovation. The `Unanticipated Events' section invites the designer to consider potential external factors from the wider world that may act on the `Status Quo' changing the foundations for the context of use of the design decision irrevocably. Though minor events, especially those directly related to a design's supply chains and sales territories, may play a more significant role in many designs' operational contexts it is important to consider major events, especially those that may be disruptive, aberrant, catastrophic or anomalous. A development of \citeauthor{aguilar1967scanning}'s EPTS \citep{aguilar1967scanning}, the `PESTLE' tool is helpful in creating macro pictures of the industry environment. `PESTLE' stands for `Political', `Economic', `Social or Socio-cultural', `Technological, `Legal' and `Environmental'. Its use here is inspired by its incorporation into \citeauthor{Taylor1990}'s futures work \citep{Taylor1990}, which specifically focuses on plausible futures. 




\subsubsection[Anticipated Consequences]{Anticipated Consequences}\label{Anticipated Consequences}

The 
`Anticipated Consequences' section focuses on the designs \textit{In use} and invites the designer to consider how proximity relates consequences directly, or indirectly, to \textit{Other Organisations}, \textit{Non-Users} and \textit{Users}, as well as the \textit{Organisation} itself. \citeauthor{rogers2010diffusion}'s diffusion theory \citep[p.~60]{rogers2010diffusion} underpins the potential consequence space. This is intended to invite consideration as to how expected, acceptable or probable a consequence of the design decision in use might be. We highlight the \textit{Users} attitude to use and the position in the adoption cycle as key points of interest. The range of designers' attitudes is given, to support a consideration of user impact and perspective, and is an extension of the spectrum presented in \citep{darby_thesis_2023}. The adoption cycle \citep{rogers2010diffusion} is used to help designers to explore  the different impacts that the point of adoption may have on usage, especially as that intersects with user attitude. The significance of this is illustrated by DJI's seeming ambivalence to the (mis)use of their drones in the war in Ukraine, see \ref{Ukraine}.

\subsubsection[Unanticipated Consequences]{Unanticipated Consequences}\label{Unanticipated Consequences}
The `Unanticipated Consequences' section focuses on the \textit{Organisation} and the planning of a design decision. Following \citeauthor{sveiby2009unintended} we draw on \citeauthor{merton1936unanticipated} to highlight the `factors that limit an actor’s possibility to anticipate both direct and indirect consequences' \citep{sveiby2009unintended}. We focus on \textit{assessment errors}, \textit{myopia} and \textit{fundamentalism}. 

\citeauthor{sveiby2009unintended} highlight \textit{myopia} as the most common limiting factor followed by \textit{assessment errors} and \textit{fundamentalism} \citep[]{sveiby2009unintended}. 
To begin to consider the `Fundamentalism' marker we must address the values that sit alongside it. \citeauthor{Schwartz} defines values as `the criteria people use to select and justify actions and to evaluate people (including the self) and events' \citep{Schwartz}. \citeauthor{Sutcliffe} introduce a values taxonomy relevant to software development \citep{Sutcliffe}. Mapping the \citeauthor{Sutcliffe} values taxonomy \citep{Sutcliffe} onto the more familiar \citeauthor{Schwartz} values framework \citep{Schwartz} demonstrated the advantage of the reduced values taxonomy as being more readily usable as part of the conceptual framework (see, section  \ref{AirTag}, our discussion of the Apple AirTag). 

As myopia occurs when the desirability of beneficial consequences becomes the overriding focus of attention it too is informed by values. Had a different set of values been embodied in the design of Horizon, see \ref{Horizon}, might the Post Office's myopic refusal to accept that bugs in the software could be responsible for accounting errors have been averted?

\subsubsection[Speculative Strategies]{Speculative Strategies}\label{Speculative Strategies}

Speculation about the future is not untethered imagination. Speculative, like prospective, design practices work within the laws of physics and focus on plausible and probable futures \cite{dunne2013} and both, while differently orientated, are underpinned by the concept of prefiguration, the need to imagine beforehand \citep{pierce2021}. To support engagement with the conceptual framework five speculative strategies drawn from \citeauthor{pierce2021}'s observation's on frictional tendencies \citep{pierce2021} are presented, alongside a cognitive heuristic, the SCAMPER brainstorming tool, \citep{eberle}, see fig \ref{fig:Consequences and Futures model}.


\citeauthor{pierce2021}'s frictional tendencies; analogical, divergent, oppositional, accelerational and counterfactual, were originally conceived as `a tool for teaching design students how to concretely compose and construct speculative, critical, and conceptual designs' \citep{pierce2021}. These were more-accessibly termed 
\textit{Associate}, 
\textit{Change tack}, 
\textit{Push back}, 
\textit{Speed the future}, 
and \textit{Swap the past}, respectively.  They are common creative design strategies seen in \citep{maidenRE2010}, \citep{brown2008}, \citep{dunne2013}, and \citep{Bleecker2009}. While SCAMPER, an acronym for Substitute, Combine, Amplify, Modify/Magnify/Minify, Put to other uses, Elaborate/Eleiminate and Rearrange/Reverse, is a tool for the development of the imagination. It is commonly used in brainstorming sessions.



NB: We do not claim that the conceptual framework is a tool for anticipation or analysis. Rather, it provides a speculative jumping off point for exploration. Tracing various pathways across the \textit{Consequences and Futures Model}, 
we aim to produce broadly plausible futures 
formed in productive tension with our understandings of probable futures.

\subsection[Workshopping the conceptual framework]{Workshopping the conceptual framework}\label{Findings3}

Across three design workshops, two with industry professionals (W1 \& W2) and another with academics (W3), the research team broadly aimed to investigate how exploratory methods from speculative design might be employed to usefully address over-the-horizon change for the purposes of understanding requirements and informing design decisions. The workshops aimed to define key properties around Digital Twins (W1 \& W2) and Search Engines and transparency (W3) through an exploration of the problem space. The individual workshops were originally structured in four parts; groundwork, speculation, diegetic prototyping and discussion.

The groundwork, based on section \ref{Status Quo} of the framework, aimed to draw out working definitions and their associated technologies, as well as brief descriptions of any current projects. Thereafter explorations of the values underpinning that work, the analogies at play, applicable trends and significant broader factors in the world were made. The participants were introduced to the speculative strategies, see \ref{Speculative Strategies} and were then asked to work at-speed developing potential futures with each of the strategies. This was followed by extended futures exercises that introduced unanticipated events, section \ref{Unanticipated Events}, as well as unanticipated consequences, section \ref{Unanticipated Consequences}, and anticipated consequences, section \ref{Anticipated Consequences}, into the mix. The next step was to have been the development of diegetic prototypes. However in W1 we drew that part of the workshop to a close early to allow time for discussion. In W2 \& W3 we followed the pattern established in W1; groundwork, speculation and discussion. 

\subsubsection[Industry practitioners Workshops]{Industry practitioners Workshops}\label{Industry practitioners Workshops}

The participants of the first two workshops (WS1: n=8, M=7, F=1; WS2: n=5, M=4, F=1) were members of a Research and Development (R\&D) team inside our industry partner. The participants were all computer scientists educated to at least first degree level, and ranging from fresh graduates to managers with many years' experience. 
Through discussion with the R\&D team leader, Digital Twin (DT) technology was selected as the focus for the workshops. DTs had emerged as a technology of interest to several of the company's business units and the R\&D team was responsible for developing a consistent understanding of DTs and evaluating their utility to the company. From our perspective, DTs was a good choice for the workshops because while DT is in some ways a new name for an old idea, a number of recent technological developments (in IoT, data science and other areas) are widening their potential utility across a range of application areas. There is still much debate, both globally and within our industry partner, about what constitutes a DT and the team leader hoped that our workshops would help the team arrive at a consistent understanding of DTs.


\begin{figure}[htbp!] \centering\setlength{\fboxrule}{0.25pt} \setlength{\fboxsep}{0pt}\fbox{\includegraphics[height=.5\textwidth]{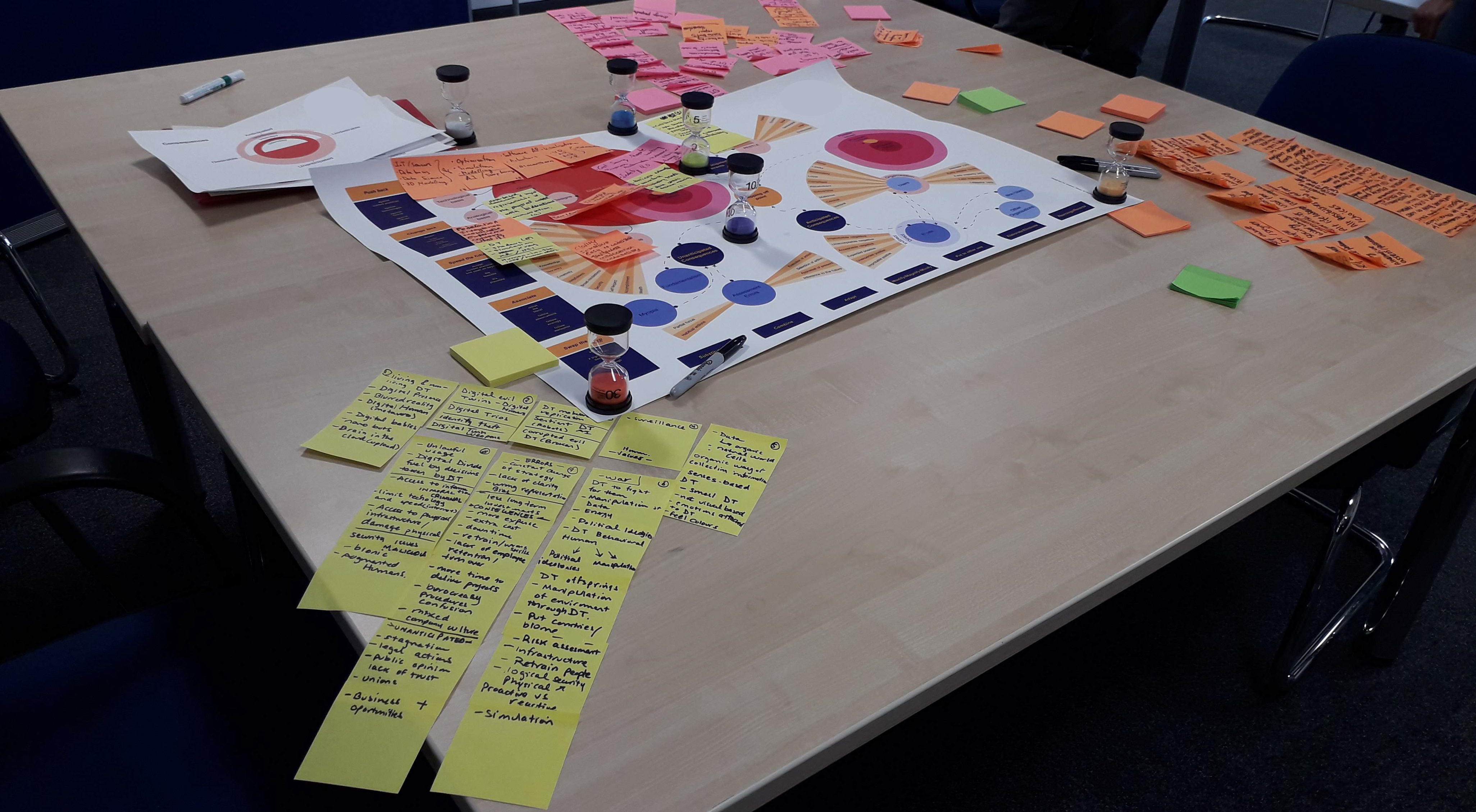}}\caption[W1 Workshop]{W1 Workshop} \label{fig:Workshop}\end{figure}

The groundwork section of W1 was covered quickly and coherently, participants agreed a working definition of DT and identified a wide range of technologies underpinning DTs, many of which are themselves new and developing. They described current DT projects, and identified internal threats to the development of future DT applications, including a lack of resources, investment, interest, and long-term direction at management level. When participants responded to our values query about their current projects they articulated the corporate values the company espouses and aspires to rather than offering a critical assessment. Also, when asked to describe the analogies and metaphors at play in their DT work the responses showed the participants' uncertainty, with abstract concepts, questions, and corporate visions offered up alongside analogies and metaphors. Responding to questions about the trends and external factors exerting an influence on future DT applications they were more certain. 


Responding to the model, W1 participants framed their understanding of DTs by developing a large speculative framework of extrapolations, predictions, and potentialities. A significant number of ideas were generated. They responded to the speculative strategies well, favouring analogical, accelerational and then oppositional strategies. They found counterfactuals hard to develop in the time available and the introduction to  deviational and divergent was rushed, which may explain fewer ideas being generated using that strategy. Throughout W1 participants were most comfortable discussing the future drawing on shared cultural references from popular TV and film, however within their work they consistently debate using high-level abstract concepts. This may explain some resistance to making any articulation of speculations concrete. Just one rough sketch of a PowerPoint slide was made in response to significant prompting from the facilitator. And the post-it note documentation was littered with one or two word encapsulations rather than more storied description. 

Participants from W1 were positive about having the space to think differently about the DTs and being encouraged to think beyond the enterprise. They also enjoyed the explicit prompting to think about what might go wrong and what the future might look like.
The observer noted that `It must have been tiring but everyone appeared to enjoy it and everyone contributed. The degree of creativity and the participants’ openness to speculation was impressive.' Interestingly, participants felt that greater diversity in the participant group might have produced different, better or more creative results. However, they were not specific as to the nature of that diversity. Addressing the potential social consequences of innovation is not a common approach within engineering teams and this fact was reflected in the R\&D team. Although the team was alert to social change and the fact that tech innovation plays a significant role in this, the team broadly viewed technological innovation as neutral, in line with instrumental theories of technology.

At the end of W1, participants identified a need to develop a more nuanced understanding of thresholds between humans and physical and virtual realities as a result of Digital Twin technologies. This informed discussion on the focus of W2, once we had established the company had had a positive experience and wanted to continue working with us.
We were asked to pivot away from a focus on unanticipated consequence towards establishing a more robust working definition of DT. The research team agreed that this would be possible building on the W1 and using the framework and particularly the speculative strategies to guide our exploration. At the outset of W2, the participants re-familiarised themselves with the post-its that documented W1. They sorted the material to establish useful ideas to define DTs.


W2 participants returned to their familiar abstract high level debate and as they formed articulations of DT applications and definitions, the research team interjected with different speculative strategies to challenge their assumptions and push them to ground their thinking in words or images. Also, drawing on aspects of the model to make specific challenges. After significant prompting they produced architectural sketches offering a high-level conceptualisation of a DT system. W1 participants navigated the speculative strategies and the model with a convergent intent. Operating in a prospective mode the industry practitioners came to a specific understanding of DTs' potential in their workplace. The process generated a new working definition of DT that was set in opposition to the original and widely accepted working definition and derived from the specifics of the company context. 


Following the W2 workshop, the participant group provided written feedback. They were broadly positive about the workshops, and pleased with the result, but were critical of the process. They thought it could be improved as it had iterated too many times without, they felt, significant additional benefit.  

\subsubsection[Academic Workshop]{Academic Workshop}\label{Academic Workshop}

The participants of the third workshop (WS3: n=5, M=3, F=2) were academics with many years' experience and postgraduates within a department of computer science educated to at least second degree level.
The research team selected Search Engines (SE) and transparency as the focus for the workshop. SEs are a well established technology that has been in constant development since before the World Wide Web. The technology's familiarity as a widely known ubiquitous development and its continued renewal made it an appropriate subject for our participant study. SEs are subject to varied development across different territories, particularly in relation to issues of transparency.



The groundwork section of W3 saw participants respond under time pressure to the topic area of Search Engines and transparency by speculating on SEs characteristics and underpinning technologies. From this it became clear that the topic area was not as familiar as we had expected. The participant responses were circumscribed by a lack of topic knowledge. However, with additional input from the facilitator (prompts to think about technologies, values, analogies, trends and factors in the wider world and how they affect SEs), an adequate working understanding of the status quo as it relates to SES was formed.


The participants responded well to the time-pressured steps leading them through the speculatative strategies, beginning with \emph{Speeding  the future} and ending with \emph{Swap the past}. This led to a flurry of creative speculations that touched on (e.g.) accuracy vs. joy of result, region-specific search results, anthropomorphised SEs, hedonistic SEs, more/less explicit provenance of results. Applying PESTLE, the participants speculated on (e.g.) explicitly costing searches in terms of their environmental impact, legal constraints on deepfakes, requirements for equality of access, etc. The participants then shifted focus to unanticipated consequences: (e.g.) more state control c.f. Weibo, shifts in users' values and preferences (attitudes to ads, value of ease of access to information), the emergence of purposely malign search engines c.f. the dark web. W3 participants navigated the model with a divergent intent. The participants, operating in an exploratory mode, came to a broad understanding of a class of technologies through its impacts. They had little time and therefore impetus to converge around a particular design or set of designs and to draw out the relation between the probable futures they described and the plausible futures they had posited.

The participants reported that they enjoyed the format of the workshop. For the postgraduate  students particularly, the idea of speculation as a means to reason about technology was novel and rewarding. Clearly, the W3 participants were under no obligation to demonstrate benefit to an employer. This was in contrast to the practitioners in W1 and W2, who, if not necessarily expected to demonstrate explicit benefit to our industry partner, did have at least a tacit expectation that the time taken out of their day-to-day work had not been wasted.





\section[Discussion]{Discussion}\label{Discussion}

\subsection[Skin in the game]{Skin in the game}\label{Discussion1}

Discussion of consequence and futures share key dimensions, the predictability and desirability of consequence as discussed in communication studies \citep[p.~60]{rogers2010diffusion} are articulated in the synonymous terms of probability and preferability in futures and speculative design discourse \citep{voros2001,dunne2013}. However, consequences are recognised as being proximal to a design decision \citep[p.~60]{rogers2010diffusion}, and are considered in relation to their potential realisation in the present. While, futures are understood as being by their nature more distal and as such they are considered in relation to their potential realisation at indeterminate points in time. Here, import correlates with immediacy of impact. Futures, even near-futures, reveal themselves slowly. After all, they are the ever-emergent result of an uncountable number of decisive actions. Pragmatically, and understandably, practitioners keep the scope of their design decisions tight and their consideration of futures even tighter. As such, the R\&D team were understandably unused to considering the potential social impact of their design decisions. While the academics, not having actual design decisions to make, found it easier to engage with futures exploration.

\subsection[Focus on unanticipated consequences]{Focus on unanticipated consequences} \label{Discussion1a}

Within technological innovation process consequences are intentionally marshalled to be as predictable, desirable and direct as possible, with any negative impacts mitigated against in development, and through supported release, using agile feedback loops. As case studies \ref{Horizon} and \ref{Ukraine} demonstrate companies sometimes fall short of this ideal with devastating social consequences. However, consequences do not have to be devastating to be consequential to the formation of plausible futures. Nor are they, necessarily, a direct result of the companies' actions, the cause and effect relation implied by unanticipated consequences is messy and muddy. The conceptual framework aims to focus attention on the territory of the long consequence, the space before the visions of slow, unfolding, remote futures, yet beyond the immediate desirable consequence of our design decisions to enable better organisational understanding of the future as a part of the design space of the present. As W1's call for greater diversity in the workshop  process attests, it is not only the chronological distance from a present that defines a future, it is the viewer's belief in a futures' possibility and probability, which is itself informed by their positionality. Any perception as to its desirability is informed by the viewers' ability to conceive its impacts on their own, and others', future lives. In this way acts of anticipation are at times perspectival.

Across all of the workshops, the various oscillations presented a significant, yet enjoyable, challenge for participants. The iterative cycle was an intentional part of the process, however, some participants questioned the approach taken with W1 \& W2, reportedly feeling they were retreading territory. This response seems almost inevitable when growing your own  unanticipated consequences from a single root stock. However, W3 participants were more positive with one going so far as to request the model for further study. For them the value resided almost tacitly in the act of determining the design, their perceived benefit was more experiential.


\subsection[Improving the model and workshop facilitation]{Improving the model and workshop facilitation}\label{Discussion2}

The conceptual framework invites the simultaneous consideration of current design decisions, informed by the status quo, in relation to a triad of contexts; the current, and recent, context of an instance of organisational technology development, the future context of the design's users-in-action, and the overarching future context of an ever-changing external world. On reflection, the original \textit{Consequences and Futures Model} contained some poorly articulated section headings. Moving forward elements of the diagram require renaming to better explain the relations in the conceptual framework. 
In accessible terms these might be, the `Planned Design Action' to give focus on the technological instance under consideration, the `Lead Organisation' to highlight the role of the organisation in the development, the `Design-in-Action' to place attention on the operational effects on both users and non-users, and the `Ever-changing World' to demonstrate the potential of uncontrollable events to impact the design decision outcomes. Communication in around consequence, futures and speculative approaches is difficult, especially when rushed by time constraints. We recognised this and provided a glossary of key terms in advance and visualised the conceptual framework to aid comprehension. We also prepared introductions for the activities and planed the facilitation. However, some inaccessibly named elements remained and some explanations could have been better. Effective communication remains a key challenge to creating a more readily usable model.

Time is a precious commodity and accessing industry practitioners was especially difficult. Our first conception of the workshop design was heavily modified to run at two hours instead of 1.5 days due to this very reason. With limited access to participants, the pacing of content was stressful at times with some becoming visibly tired by the demands of the workshop. A number of the activities demand time. For example, participants' familiarity and facility with analogies and metaphor is likely to be variable, and allowing time to introduce these literary devices properly is essential to support full engagement with the framework. Convincing organisational gatekeepers of the value of the programme and the demands of the series of exercises becomes an essential project task for the facilitator/researcher.




\section[Limitations and Future work]{Limitations and Future work}\label{Limitations and Future work}


Feedback from the workshop participants and our own observations of how the participants' understanding evolved provides evidence that our model can help stakeholders gain a wider understanding of a technology and its impact. This has been most strongly demonstrated where our participants were operating in exploratory mode, seeking to better understand a technology and its potential for impact.

For people working in prospective mode, as were our industry participants, there is much still to do. Our industry participants had an R\&D rather than a product development role. Using the model, our participants proved adept at using the model to find commonality among the multiple and sometimes poorly-aligned company-wide conceptualisations of digital twin technologies and thus postulating new product features.  The model stimulated the team's creativeness in a way that was possibly useful for the company. However, when discussion led the participants into potentially negative impacts of envisioned developments of digital twins, e.g. deskilling, they didn't pursue the line of reasoning that led them there, instead tending to gloss over the negatives. This limitation may be a reflection of the fact that digital twins were already a de-facto technology within the company, even if their maturity of adoption varied considerable across the company's different divisions and business functions. We postulate that had the team been tasked with evaluating an emerging technology, not yet adopted by the company, there might have been more appetite for a more balanced investigation of the potential for both  +ve and -ve  impacts.

Ultimately, our goal is that insights into technology impacts should feed into design decisions. For this to happen, a method is needed that stakeholders see as clearly beneficial; that can help designers avoid unanticipated or unintended negative consequences and which can be integrated with existing requirements and design practices. To better understand the extent to which our model represents an advance towards this goal, we will need to transform such insights into product requirements in a live project and trace how requirements are handled throughout a full project life-cycle. 




\section[Conclusion]{Conclusion}\label{Conclusion}

While we do not claim to have developed a `reliable methods for anticipating undesirable consequences of innovation' \citep{sveiby2009unintended} we have made progress towards a useful method. The study demonstrated potential for the conceptual framework to be used as a tool with implications for research and practice.

The conceptual framework has a potential role in the training of technologists, particularly computer scientists and software engineers. We believe recent evidence suggests that developers are poorly equipped to consider the consequences of design decisions and insufficiently sensitised to the potential for their actions to impact beyond a narrow stakeholder group.

Most CS and SE degree programmes are focused on training students to solve problems through software. Problem understanding typically forms a small part of the curriculum, perhaps in a requirements engineering module. The potential of our work is to help reorient the student focus from solution finding to problem understanding and, beyond that, to problem discovery. Here, developers learn the potential benefits of exploring the problem space beyond that defined in (e.g.) a customer brief or market opportunity to project a little into the future and consider what assumptions might prove to be poorly-founded and to think through the impacts of this happening.



\section{Acknowledgements} This paper results from the Twenty20Insight project funded by the UK’s Engineering and Physical Sciences Research Council (Grant Ref. EP/T017627/1). The research study had ethical approval from the Science Faculty Ethics Committee at Durham University (Reference: COMP-2022-05-04T16\_15\_46-hdcl45). Many thanks to our participants for generously sharing their time, expertise and enthusiasm. 



\bibliographystyle{ACM-Reference-Format}
\bibliography{andyrefs}
\end{document}